\documentclass[12pt]{article} 
\usepackage{a4} 
\usepackage{epsfig}
\usepackage{amssymb} 
\usepackage{amsmath} 
\usepackage{amsfonts}

\def\vectau{{\pmb{\tau}}}
\def\vecpi{{\pmb{\pi}}}
\def\vecomega{{\pmb{\omega}}}

\newtheorem {theorem}{Theorem}[section]
\newtheorem {lemma}[theorem]{Lemma}
\newtheorem {corollary}[theorem]{Corollary}

\begin{document}
\title{
  \vskip 15pt
  {\bf \Large \bf Homotopy of Rational Maps and the
    Quantization of Skyrmions}
  \vskip 10pt}
\author{
Steffen Krusch\thanks{S.Krusch@maths.leeds.ac.uk } \\[5pt]
{\normalsize {\sl Department of Pure Mathematics }}\\
{\normalsize {\sl School of Mathematics, University of Leeds,
Leeds LS2 9JT}}\\
{\normalsize {\sl England}} 
}

\date{October 29, 2002}

\maketitle

\begin{abstract}
The Skyrme model is a classical field theory which models the strong 
interaction between atomic nuclei. It has to be quantized in order to 
compare it to nuclear physics. When the Skyrme model is semi-classically 
quantized it is important to take the Finkelstein-Rubinstein constraints 
into account. The aim of this paper is to show how to calculate these FR 
constraints directly from the rational map ansatz using basic homotopy 
theory. We then apply this construction in order to quantize the Skyrme 
model in the simplest approximation, the zero mode quantization. This is 
carried out for up to 22 nucleons, and the results are compared to 
experiment.
\end{abstract}

\vspace{2cm} 
\centerline{PACS-number: 12.39.Dc}
\vspace{2cm}
\begin{tt}
\pageref{lastref} pages, 2 tables
\end{tt}

\newpage

\section{Introduction}

The Skyrme model is a classical model for the strong interaction
between atomic nuclei, \cite{Skyrme:1961vq}.  Any field configuration
in this model is labelled by a topological winding number which can be
interpreted as the baryon number $B$. In order to compare the Skyrme
model to experiments it has to be quantized.  Since a canonical
quantization of the model is not possible because it is
non-renormalizable as a field theory, the model can only be quantized
approximately. 

In \cite{Adkins:1983ya, Adkins:1984hy} Adkins {\it et al.} quantized
the rotational and translational zero modes of the $B=1$ Skyrmion and
obtained good agreement with experiment. A subtle point is that the
Skyrme model, as a classical field theory, allows for quantizing
Skyrmions as fermions. This is because the configuration space is not
simply connected \cite{Finkelstein:1968hy}. For the $SU(2)$ Skyrme
model there is a choice whether to quantize a Skyrmion as a boson or a
fermion when $B$ is odd. Yet, it has to be quantized as a boson if $B$
is even, \cite{Giulini:1993gd}.  Witten showed in
Refs. \cite{Witten:1983tw, Witten:1983tx} that in the $SU(3)$ model
the $B=1$ Skyrmion has to be quantized as fermion if the number of
colours is odd, yet no such restriction applies to the $SU(2)$ Skyrme
model. 

In order to quantize Skyrmions with higher baryon number it is
important to understand the classical solutions.  Well-separated
Skyrmions can be described reasonably well by the product ansatz,
however when Skyrmions coalesce this ansatz fails. In fact, 
the static solutions of the Skyrme model have a surprisingly rich 
structure.  
Braaten {\it et al.} showed
in Ref. \cite{Braaten:1990rg} 
that the $B=2$ Skyrmion is a torus, the
$B=3$ Skyrmion is a tetrahedron and the $B=4$ Skyrmion is a
cube. Battye and Sutcliffe calculated the Skyrmions up to $B=9$,
\cite{Battye:1997qq} and later to $B=22$, \cite{Battye:2000se}, and
also found shell-like structures with discrete symmetries. 

The $B=2$ Skyrmion with toroidal symmetry was quantized in Refs.
\cite{Braaten:1988cc,Verbaarschot:1987au} using the zero mode
quantization. Later, the approximation was improved by taking massive
modes into account, \cite{Leese:1995hb}.  The $B=3$ Skyrmion was first
quantized in \cite{Carson:1991yv}. Then the cubically symmetric $B=4$
Skyrmion was quantized in \cite{Walhout:1992gr}. Irwin has performed a
zero mode quantization for $B=4$ to $B=9$, \cite{Irwin:1998bs}, where
the monopole moduli space was used as an approximation for the Skyrmion
moduli space. For all the even baryon numbers he found that the
predicted quantum numbers of the ground state agreed with nuclear
physics.  Yet, his findings disagreed with experiments for $B=5$, $7$
and $9$. 

The aim of this paper is to construct the Finkelstein-Rubinstein
constraints directly from the rational map ansatz,
\cite{Houghton:1998kg}. With this ansatz it is possible to understand
the symmetries of Skyrmions. Effectively, it gives a finite dimensional
approximation to the configuration space which is more tractable than
the infinite dimensional configuration space but still retains 
important topological properties.  Moreover, the ansatz plays a
major role in constructing Skyrmions.  To date, both the numerical
configurations and the relevant rational maps are known up to $B=22$,
\cite{Battye:2000se, Battye:2001qn}. In this paper we apply these 
results to calculate the quantum ground states up to $B = 22$ using zero 
mode quantization. 
  
This paper is organised as follows. In Sect. \ref{SkyrmeModel} we briefly
review the Skyrme model and discuss its topology in some detail. The
topology, in particular the fundamental group of configuration space, is
very important for understanding the Finkelstein-Rubinstein constraints
for quantizing scalar fields as fermions \cite{Finkelstein:1968hy}. We
describe this construction and its implications on zero mode quantization.
In Sect. \ref{Rational} we recall the rational map ansatz 
\cite{Houghton:1998kg} and show that it can be viewed as a suspension.
This enables us to calculate the fundamental group of configuration space
directly from rational maps. Applying some theorems on the fundamental
group of rational maps we prove a simple formula to calculate the
homotopy class of a loop generated by a combined rotation and isorotation.
In Sect. \ref{Results} we discuss how to use group theory to find the 
ground states in the Skyrme model and present the results of our 
calculations. We end with a conclusion.

\section{The Topology of the Skyrme Model}
\label{SkyrmeModel}

The Skyrme model is a classical field theory of mesons.  The basic
field is the $SU(2)$ valued field $U({\bf x},t)$ where ${\bf x} \in
{\mathbb R}^3$.  The static solutions of the Skyrme model can be
derived by varying the following energy \cite{Houghton:1998kg}: 
\begin{equation}
\label{Estandard}
E = \int \left( -\tfrac{1}{2} {\rm Tr} \left( R_i R_i \right) -
\tfrac{1}{16} {\rm Tr} \left( [R_i, R_j][R_i,R_j] \right) \right) {\rm
d}^3 {\bf x}, 
\end{equation}
where $R_i = (\partial_i U) U^\dagger$ is a right invariant
${\mathfrak{su}}(2)$ valued current.  A static solution of the
variational equations could be a saddle point. Only solutions
which minimise the energy are called Skyrmions.  In order to
have finite energy the Skyrme fields have to take a constant value,
$U(|{\bf x}| = \infty) = 1$, at infinity.\footnote{The value of $U$ at 
infinity is fixed by assuming there is a (infinitesimal) pion mass term.}
Therefore, the space of Skyrme configurations consists of all 
maps $U: 
{\mathbb R}^3 \to S^3$ with boundary condition $U(|{\bf x}| = \infty)
= 1$ which effectively compactifies ${\mathbb R}^3$ to $S^3$. 
Such  maps can be characterised by their degree which
is an element of the third  homotopy group $\pi_3(S^3) \cong
{\mathbb Z}$. We will call the space of Skyrme configurations $Q^*$ in 
order to emphasise that the maps are based, {\it i.e.}  the point 
``$\infty$'' is mapped to
$1$. The configuration space $Q^*$ is split into
connected components $Q_B^*$ which are the homotopy classes of
$\pi_3(S^3)$. The fact that $Q_B^*$ is connected follows from a
famous theorem of Hopf ({\it e.g.} \cite{Hatcher:2002}).
Furthermore, the energy of configurations in $Q^*_B$ is bounded below by the 
topological charge: $E \ge 12 \pi^2 B$, \cite{Faddeev:1976pg}.

\subsection{Finkelstein-Rubinstein Constraints}
\label{FRconstraints}

In the following we describe an idea of Finkelstein and Rubinstein, how to 
quantize a scalar field theory and obtain fermions. Quantization usually
implies replacing the classical configuration space by (wave) functions on 
configuration space. 
Finkelstein and Rubinstein argued that if the configuration space 
$Q^*$ is not simply connected, then the wave functions have to be defined 
not on configuration space $Q^*$, but on the covering space of 
configuration space $CQ^*$. To simplify matters, we assume that baryons
are conserved in our theory, that is baryons cannot decay. This means that 
the wave function has to be non vanishing only on one component, otherwise 
there could be transitions between different sectors. Therefore, we impose 
the (superselection) rule that the wave functions are defined on $CQ^*_B$.

In order to have fermionic quantization, a rotation of a wave function 
$\psi$ by $2 \pi$ has to result in $-\psi$. However, the $SO(3)$ action  
is not well defined on $CQ^*_B$. In order to define this group action one 
has to keep track of the component of the covering space and this is the 
origin of the FR constraints.

Finkelstein showed in Ref. \cite{Finkelstein:1966ft} that the
fundamental group $\pi_1(Q_0^*)$ is isomorphic to $\pi_4(S^3)$. 
It is a standard  result of algebraic topology that $\pi_4(S^3) \cong
{\mathbb Z}_2$, {\it e.g.} \cite{Hatcher:2002}. Furthermore, Whitehead
proved the following theorem in Ref. \cite{Whitehead:1946}.
\begin{theorem}[Whitehead]
\label{Whitehead}
Let $F^p(X,x_0)$ be based maps $f: S^p \to X$ such that $f(1) = x_0$.
Then the connected components of $F^p(X,x_0)$ are homotopy equivalent. 
\hfill $\square$
\end{theorem}
This implies in particular that the fundamental groups $\pi_1(Q_{B_1}^*)$
and $\pi_1(Q_{B_2}^*)$ for different components  are isomorphic.
It follows that\footnote{An alternative proof was given in Ref. 
\cite{Giulini:1993gd} which relies on the fact that the target space of  
the Skyrme model is a group, namely $SU(2)$.}
\begin{equation}
\pi_1(Q_B^*) \cong \pi_4(S^3) \cong {\mathbb Z}_2.
\end{equation}

The covering space $CQ_B^*$ can now be defined as the space of all 
paths starting at a fixed point $q_0 \in Q_B^*$ modulo 
homotopy. Elements in $CQ_B^*$ will be denoted by $[q]$. Let $[a]$ 
be the generator of the fundamental group of $Q_B^*$. 
Then we can define the 
path $[q \cdot a]$, where $[a]$ is a loop, starting at $q$. 
This is well defined because $Q_B^*$ is connected so that the loop is 
independent of the base point $q_0$ (up to isomorphism). 
$[q \cdot a]$ and $[q]$ are different points in $CQ_B^*$ which 
project  to the same point $q \in Q_B^*$.
 As $\pi_1(Q_B^*) = {\mathbb Z}_2$, $[a^2]$ is a trivial loop, 
therefore, 
\begin{equation}
\label{a^2}
[q\cdot a \cdot a] = [q].  
\end{equation}

Define a wave function 
\begin{equation}
\psi: CQ^* \times {\mathbb R} \to {\mathbb C}: 
([q],t) \mapsto \psi([q],t) 
\end{equation}
as an element of a formal Hilbert space ${\cal H}$, such that 
$\psi$ is square integrable and normalised:
\begin{equation}
\int |\psi([q],t)|^2 {\rm d} [q] = 1.  
\end{equation}
Define $L_a \psi([q],t) = \psi([q\cdot a], t)$. 
Conventionally, a wave function is determined by configuration 
space, modulo a phase. Then
\begin{equation}
L_a \psi([q],t) = {\rm e}^{i \phi} \psi([q],t),
\end{equation}
and equation (\ref{a^2}) implies that ${\rm e}^{2i \phi} = 1$.
If the phase is trivial then we can define the function on 
$Q_B^*$. In physicists' language, this corresponds to a theory which 
only contains bosons. Therefore, we impose a nontrivial phase
\begin{equation}
\label{noncontractible}
L_a \psi([q],t) = - \psi([q],t).
\end{equation}
We now impose rotational and isorotational symmetry on the wave 
function as follows. We insist that
the wave function $\psi$ vanishes on the path 
$[q]$ unless the corresponding configuration $q$ is  
invariant under a rotation through angle $\alpha$ around axis 
${\bf n}$ followed by an isorotation through angle $\beta$ around 
axis ${\bf N}$. Then equation (\ref{noncontractible}) implies the 
following constraint on the wave function:
\begin{equation}
\label{FR}
{\rm e}^{-i \alpha {\bf n} \cdot {\bf J}}
{\rm e}^{-i \beta {\bf N} \cdot {\bf I}}
\psi([q],t) 
= \left\{ 
\begin{array}{r l}
\psi([q],t) & {\rm if\ the\ induced\ loop\ is\ 
contractible,} \\
- \psi([q],t)      & {\rm otherwise.} 
\end{array}
\right.
\end{equation}
Here ${\bf J}$ and ${\bf I}$ are the spin and isospin operators 
respectively, with quantum numbers $J$ and $I$.
This imposes a constraint on the allowed quantum states, namely 
equation (\ref{FR}) can only be satisfied for suitable values of 
$I$ and $J$. 
Particularly important are 
rotations or isorotations by $2 \pi$ because they are always a 
symmetry of the system. When a $2 \pi$ rotation gives rise to a 
nontrivial loop then this corresponds to half integer spin.
It is worth noting that Finkelstein and Rubinstein 
proved a connection between spin and statistics in this setting. 
Namely, $Q_B^*$ admits half integer spin if and only if it admits odd 
exchange statistics.

\subsection{Zero Mode Quantization}
\label{Zero-mode}

The Skyrme model is not renormalizable as a field theory. The usual 
approach is to quantize it semi-classically. The key idea is the 
following. The classical dynamics of (slow-moving) solitons can often be 
described by geodesic motion on the moduli space of static solutions 
\cite{Manton:1982mp}. This has 
been shown for monopoles in \cite{Gibbons:1986df}, where this idea 
was also used to quantize monopoles. For an overview of exact results 
about the geodesic approximation for vortices, see {\it e.g.} 
\cite{Manton:2002wb} and references therein. 
Contrary to monopoles and vortices the minimum energy configuration of a
given sector is found to be unique up to the action of the symmetry
group, namely translations in space and rotations 
in space and target space. For $B=1$ the Skyrmion has spherical 
symmetry so that the symmetry orbit is 6 dimensional. For $B=2$ there is 
axial symmetry and the orbit is $8$ dimensional, whereas for higher 
baryon number there are only discrete symmetries and the orbit is 9 
dimensional. 
Since the formal Hilbert space ${\cal H}$ of the previous section is 
difficult to handle in practice,  a simple approximation 
is often used in the literature, namely that the 
wave function is only non-vanishing on the static minimal energy solution
of a given sector, {\it i.e.} a Skyrmion, and its symmetry orbit.
This quantization is known as zero mode 
quantization.\footnote{Here, the word ``zero mode'' is used rather loosely 
to refer to symmetry transformations that leave the energy 
(\ref{Estandard}) invariant.}

The Skyrme Lagrangian is invariant under the Poincar\'e group of (3+1) 
dimensional space, $SO(3)$ rotations in target space and some discrete 
parity transformations, which will not be considered here. 
Similarly, the space of static solutions, that is configurations which 
minimise the energy (\ref{Estandard}), is invariant under the Euclidean 
group ${\mathbb E}_3$, isorotations and parity transformations. 
By acting with the latter symmetry group on a static Skyrmion $U_0$ 
we generate a set of new static solutions
\begin{equation}
\label{zeromodes}
U({\bf x}) = A U_0 \left(D(A^\prime) \left({\bf x}-{\bf X}\right)\right)
A^\dagger,
\end{equation}
where $A$ and $A^\prime$ are ${\rm SU}(2)$ matrices, $D(A^\prime)$ is
the  associated  ${\rm SO}(3)$ rotation and ${\bf X}$ is a vector. In
the zero mode approximation the matrices $A$ and $A^\prime$ and
the vector ${\bf X}$ are considered to be time dependent. This leads
to the following reduced Lagrangian
\begin{equation}
\label{Lred}
L = -M + \tfrac{1}{2} M {\dot X}^2 + \tfrac{1}{2} a_i U_{ij} a_j  +
\tfrac{1}{2} b_i V_{ij} b_j - a_i W_{ij} b_j,
\end{equation}
where $M$ is the classical mass of the Skyrmion, and
\begin{equation}
\begin{array}{c c c}
a_k = -i {\rm Tr} \left(\tau_k A^\dagger {\dot A} \right)  & {\rm and}
&   b_k = -i {\rm Tr} \left(\tau_k {\dot {A^\prime}} 
{A^\prime}^\dagger \right).
\end{array}
\end{equation}
The matrices $U_{ij}$, $V_{ij}$ and $W_{ij}$ are dependent on the
classical solution $U_0$ and are given by \cite{Braaten:1988cc} 
\begin{eqnarray}
\label{intU}
U_{ij} &=& - \tfrac{1}{8} \int d^3 x~ {\rm Tr} \left( T_i T_j +
\tfrac{1}{4} \Big[R_k, T_i \Big] \Big[R_k, T_j \Big] \right), \\ 
\label{intW}
W_{ij} &=&
\tfrac{1}{8} \int d^3 x~ \epsilon_{jlm} x_l~  {\rm Tr} \left( T_i R_m +
\tfrac{1}{4} \Big[R_k, T_i \Big] \Big[R_k, R_m \Big] \right),\\ 
\label{intV}
V_{ij}
&=& -\tfrac{1}{8} \int d^3 x~ \epsilon_{ilm} \epsilon_{jno} x_l x_n~
{\rm Tr} \left( R_m R_o + 
\tfrac{1}{4} \Big[R_k, R_m \Big] \Big[R_k, R_o \Big]
\right),
\end{eqnarray}
where $R_k = (\partial_k U_0) U_0^\dagger$ is the right invariant 
${\mathfrak{su}}(2)$ current which has been defined above, and 
\begin{equation}
T_i = i \Big[\frac{\tau_i}{2},U_0 \Big] U_0^\dagger
\end{equation}
is also an ${\mathfrak{su}}(2)$ current.
Lagrangian (\ref{Lred}) is no longer Lorentz invariant which is consistent 
with the fact that the moduli approximation only works for small 
velocities. For a covariant treatment for $B=1$ see for example 
\cite{Kruczenski:1997xc}.

Lagrangian (\ref{Lred}) can now be canonically quantized. 
The momenta
conjugate to $a_i$ and $b_i$ become the body-fixed spin and isospin
angular momenta $K_i$ and $L_i$ satisfying the $SU(2)$ commutation
relations, see \cite{Braaten:1988cc} for details. 
The usual space-fixed spin and isospin angular momenta
$J_i$  and $I_i$ are related to the body-fixed operators by
\begin{equation}
\begin{array}{c c c}
J_i = - D_{ij} (A^\prime) L_j, & & I_i = - D_{ij} (A) K_j.
\end{array}
\end{equation} 
The remaining nonvanishing commutation relations are
\begin{equation}
\begin{array}{c c c}
\big[L_i,A^\prime \big]  = - \frac{\tau_i}{2} A^\prime, & &
\big[
J_i,A^\prime \big]  = \frac{\tau_i}{2} A^\prime, \\ \\ \big[I_i,A
\big]  = - \frac{\tau_i}{2} A, & & \big[K_i,A \big]  =
\frac{\tau_i}{2} A.
\end{array}
\end{equation}
Therefore, ${\bf L}^2 = {\bf J}^2$ and ${\bf I}^2 = {\bf K}^2$.  A
basis for this Hilbert space is given by
\begin{equation}
\label{eigenstates}
|J,J_3,L_3\rangle \otimes |I,I_3, K_3 \rangle
\end{equation}
with $-J \le J_3, L_3 \le J$ and $-I \le I_3, K_3 \le I$. 
In this approximation, the ground
states are the states with the lowest values of $I$ and $J$ that are
compatible with the FR constraints arising from the symmetries of the
given Skyrmion.

The values of the integrals $U_{ij}$, $V_{ij}$, and $W_{ij}$ strongly 
depend on the symmetries of the Skyrmion. For tetrahedral symmetry $T$,
octahedral symmetry $O$, and icosahedral symmetry $Y$, the matrices often
only have one eigenvalue, see \cite{Irwin:1998bs} 
for a detailed discussion. In this case, the Hamiltonian is that of a 
spherical top. If there is an axis of symmetry of higher than second order 
then the matrices have at most two eigenvalues. If these eigenvalues 
are distinct then the Hamiltonian is that of a symmetric top. In both 
cases, the states in (\ref{eigenstates}) will be energy eigenstates as 
well. If the states only have dihedral $D_2$ symmetry, then the 
Hamiltonian is that of an asymmetric top and the states
(\ref{eigenstates}) are no longer energy eigenstates, see {\it e.g.}
\cite{Landau:1977}.

The integrals $U_{ij}$, $V_{ij}$, and $W_{ij}$ determine the values 
of the moment of inertia for $J$ and $I$. For $B=1$ these moments 
of inertia are equal, however, in general the moments of inertia 
are larger for rotations than for isorotations, which implies that 
for $I=J$ isorotations contribute more to the energy.

\section{The Rational Map Ansatz}
\label{Rational}

Equation $(\ref{Estandard})$ can only by solved numerically. However,
there is a good approximation, called the rational map ansatz
\cite{Houghton:1998kg}. In this ansatz Skyrme fields as maps from $S^3
\to S^3$ are given in terms of rational maps which are holomorphic
maps from $S^2 \to S^2$. 
This ansatz not only gives good approximations for 
the Skyrme configurations and in particular their symmetries. It 
also gives a good approximation to the topology of configuration 
space $Q_B^*$. We will prove the following theorem.

\begin{theorem}
\label{t:FR}
The rational map ansatz induces a surjective homomorphism from the 
fundamental group of 
based rational maps $\pi_1(Rat_B^*)$ onto the fundamental group of 
Skyrme configurations $\pi_1(Q_B^*)$. 
\end{theorem}

\subsection{Description of the Rational Map Ansatz}
\label{Ransatz}

The rational map ansatz is best derived in the geometric approach 
to the Skyrme model \cite{Manton:1987xt}. For more details see
\cite{Houghton:1998kg, Krusch:2000gb}. Here, we only state the main 
results.
First note that the angular coordinates $(\theta, \phi)$ can be
related  to the complex plane $z$ by the stereographic projection 
$z={\rm e}^{i \phi}  \tan \frac{\theta}{2}$. Then the rational map 
ansatz is given by
\begin{equation}
\label{Urational}
U({\bf x}) = \exp (i f(r)\ {\hat {\bf n}}_R(z) \cdot \vectau).
\end{equation}
Here $f(r)$ is a shape function with boundary condition $f(0) = \pi$
and  $f(\infty) = 0$, and $\tau_i$ are the Pauli matrices. The
angular  behaviour is determined by the function  $R(z)$ which is 
also the stereographic projection of an $S^2$ in target space 
$S^3$. The unit vector ${\hat {\bf n}}_R(z)$ is  
\begin{equation}
\label{normalR}
{\bf {\hat n}}_R = \frac{1}{1 + |R|^2} \left( 2 \Re(R), 2 \Im(R), 1 -
|R|^2 \right).
\end{equation}
There is one further restriction on the map $R(z)$ in that it is a
holomorphic map of degree $B$. This implies that $R(z)$ can be written
as  a ratio of two polynomials,
\begin{equation}
R(z) = \frac{p(z)}{q(z)},
\end{equation} 
where $p(z)$ and $q(z)$ are polynomials with maximal polynomial degree
$B$ which have no common factor.  Moreover, at least one of the 
polynomials has polynomial degree $B$, see {\it e.g.} 
\cite{Jarvis:1996}. We will call the class of such maps 
$Rat_B$ and will show in Sect. \ref{Suspension} 
that such maps give rise to configurations with baryon
number $B$.

In practice, the rational map ansatz is used in the following way.
Inserting (\ref{Urational}) into (\ref{Estandard}) gives 
\begin{equation}
\label{Erational}
E = 4 \pi \int  \left( {f^\prime}^2 r^2 + 2 B ({f^\prime}^2 + 1)
\sin^2 f  + {\cal I} \frac{\sin^4 f}{r^2} \right) {\rm d} r,
\end{equation}
where\footnote{${\cal I}$ is also known as the harmonic 4-energy of 
holomorphic maps $R: S^2 \to S^2$.}
\begin{equation}
{\cal I} = \frac{1}{4 \pi} \int \left( \frac{1 + |z|^2}{1 +|R|^2}
\left|\frac{{\rm d} R}{{\rm d} z} \right| \right)^4 \frac{ 2 i {\rm
d}z {\rm d} {\bar z}}{(1+|z|^2)^2}.
\end{equation}
Now, the minimum ansatz field is found by choosing polynomials $p(z)$
and  $q(z)$ which minimise ${\cal I}$ and then calculating the shape
function  $f(r)$ numerically.  These minimum energy ansatz fields have
been calculated for all the known Skyrmions and are found to have
energies that exceed the true, numerically determined minima by less
than $3 \%$. Moreover, in almost all the cases the rational map ansatz
gives the correct symmetries of the Skyrme fields
\cite{Battye:2000se}. In the following, we will describe the symmetries of 
Skyrmions, and their rational maps, in more detail.

Skyrmions can be rotated by an $SU(2)$ matrix $A^\prime$ and isorotated 
by $A$ as in equation (\ref{zeromodes}).
The matrices $A$ and $A^\prime$ induce the following 
transformations of the rational map $R(z)$.
\begin{equation}
R(z) \to {\tilde R}(z) = M_A \left( R \left(M_{A^\prime}(z) \right) 
\right).
\end{equation}
Configurations are invariant under a subgroup of these
transformations if ${\tilde R}(z) = R(z)$. 
Since the corresponding M\"obius transformations 
$M_A$ and $M_{A^\prime}$ do not agree with the canonical $SU(2)$ action, 
we derive the transformations explicitly.

In 3-dimensional space a rotation by $\theta$ around the unit vector 
$\vecomega$ is given by $\exp(- i \theta \omega_i J_i)$ where 
the (space fixed) angular momentum  operator is 
$(J_k)_{lm} = -i \epsilon_{klm}$. 
In order to describe rotations for rational maps we relate the unit 
vector ${\bf n}$ to a complex vector
\begin{equation}
V = \left(
\begin{array}{c}
v_1 \\
v_2
\end{array}
\right) \in \mathbb{C}^2
\end{equation}
via the formula
\begin{equation}
\label{defV}
n_i = \frac{V^\dagger \tau_i V}{V^\dagger V},
\end{equation}
where $\tau_i$ are the standard Pauli matrices. If $V^\dagger V =  1$ then 
this map is the famous Hopf map. Note also that because of the denominator 
in equation (\ref{defV}) this map is well defined for $V \in \mathbb{CP}^1$ 
where $[v_1,v_2] \cong [\lambda v_1, \lambda v_2]$ for $\lambda \in 
{\mathbb C}^\times$. In this case the map is one to one.

Let
\begin{equation}
R_{ij} = \exp\left(-i \theta \omega_k J_k \right)_{ij}
\end{equation}
be a rotation by $\theta$ around the unit vector $\omega_i$. Then the 
rotation of ${\bf n}$ corresponds to a $SU(2)$ rotation $A^\prime$ of $V$: 
\begin{equation}
n_i^\prime = R_{ij} n_j 
= \frac{V^\dagger {A^\prime}^\dagger \tau_i A^\prime V}{V^\dagger 
{A^\prime}^\dagger A^\prime V} 
\end{equation}
A straight forward calculation shows that $A^\prime$ is given by
\begin{equation}
A^\prime = \pm \exp\left(-i \frac{\theta}{2} \omega_k \tau_k \right).
\end{equation}
The complex number $z$ is related to ${\bf n}$ by the stereographic 
projection
\begin{equation}
z = \frac{n_1 + i n_2}{1+n_3} = \frac{v_2}{v_1},
\end{equation}
where the second equality follows from the definition (\ref{defV}) by 
direct calculation. Therefore, we obtain the transformation law 
\begin{equation}
{\tilde z} = \frac{\big( \cos \frac{\theta}{2} + i \omega_3 \sin
\frac{\theta}{2} \big) z + \big(\omega_2 - i \omega_1 \big) \sin
\frac{\theta}{2}} 
{\big(-\omega_2 - i \omega_1 \big) \sin
\frac{\theta}{2} z +\big( \cos \frac{\theta}{2} - i \omega_3 \sin
\frac{\theta}{2} \big)}.
\end{equation}

Now, let $A=\exp(-i \theta \omega_i I_i)$ be the isospin rotations by 
$\theta$ around  the unit vector $\omega_i$ and $I_i = \frac{1}{2} 
\tau_i$. Then $A$ induces a rotation $R_{ij}$ on the unit vector ${\bf 
n}_R$ in equation (\ref{normalR}). We repeat the derivation 
above and obtain
\begin{equation}
{\tilde R} = \frac{\big( \cos \frac{\theta}{2} + i \omega_3 \sin
\frac{\theta}{2} \big) R   + \big(\omega_2 - i \omega_1 \big) \sin
\frac{\theta}{2}} 
{\big(- \omega_2 - i \omega_1 \big) \sin
\frac{\theta}{2} R +\big( \cos \frac{\theta}{2} - i \omega_3 \sin
\frac{\theta}{2} \big)}.
\end{equation}
Note that rational maps can also be defined as maps $\mathbb{CP}^1 \to 
\mathbb{CP}^1$ using homogeneous coordinates $[v_1,v_2] \to [p,q]$, 
where square 
brackets imply the equivalence relation $[u,v] = [\lambda u, \lambda v]$ 
for $\lambda \in {\mathbb C}^\times$. Then $p$ and $q$ can be considered as 
homogeneous polynomials in $v_1$ and $v_2$, see \cite{Houghton:1998kg, 
Houghton:2001fe} for details.
The advantage of these coordinates is that both the 
rotation group and the isospin rotation group act linearly.

\subsection{The Rational Map Ansatz as a Suspension}
\label{Suspension}

A construction like the rational map ansatz is very common in
algebraic topology and is known as a suspension. In fact the
suspension ${\cal S} M$  of a given manifold $M$ is constructed by
considering the product space  $I \times M$, where $I$ is the interval
$I=[0,1]$, and taking the quotient with respect  to the two end points
of $I$: 
\begin{equation}
{\cal S} M = (I \times M) / (\{0\} \times M \cup \{1\} \times M).
\end{equation}
In other words, there is an equivalence relation $\sim$ which identifies 
all the points $\{0\} \times M$ to one point and also identifies all the 
points $\{1\} \times M$ to one point. 
The quotient space $(I \times M)/\sim$ is again a manifold.
One reason why this construction is so important is its 
natural compatibility with spheres:  ${\cal S} S^n = S^{n+1}$.  However, not
only spaces, but also maps can be suspended. Let $f: M \to N$ then the
suspension ${\cal S} f$ of $f$ is given by  ${\cal S} f: {\cal S} M
\to {\cal S} N$. This map is induced by $Id \otimes f: I \times M \to I \times 
N$ where $Id: I \to I$ is the identity map.
The suspension is very convenient for studying homotopy 
group as the following theorem suggests, see 
\cite[Corollary 4.24]{Hatcher:2002}.
\begin{theorem}[Freudenthal suspension theorem]
\label{t:suspension}
The suspension map 
\linebreak[4]
\mbox{$\pi_i(S^n) \to \pi_{i+1}(S^{n+1})$} is an
isomorphism for $i < 2n-1$ and a surjection for $i= 2n - 1$. \hfill 
$\square$
\end{theorem}
It is easy to verify that the rational map ansatz is a suspension. 
Both domain and codomain are given by Cartesian 
products of an interval and a two-sphere. The fact that the radial 
coordinate $r \in [0,\infty)$ can be considered as a closed interval 
relies on the boundary conditions. Moreover, at each of the end 
points the map collapses to a point. So domain and codomain are 
suspensions of $S^2$ and the rational map ansatz suspends holomorphic maps 
from $S^2 \to S^2$.  

An immediate consequence of theorem \ref{t:suspension} is that rational
maps of degree $B$ induce Skyrme configurations of degree $B$ which proves our 
claim in Sect. \ref{Ransatz}. In the
following we prove theorem \ref{t:FR}. This theorem can be used to
calculate the FR phase directly from rational maps.
Denote the space of based rational maps $Rat_B^*$ by
\begin{equation}
Rat_B^* = \{R \in Rat_B: R(\infty) = 1 \}.
\end{equation}
In order to apply the Freudenthal suspension theorem \ref{t:suspension} we 
need to consider more general maps between two-spheres.
Let $M_B$ denote continuous maps from $S^2 \to S^2$ of degree 
$B$, and  $M_B^*$ the space of based maps in $M_B$. The following 
theorem by Segal, \cite{Segal:1979}, implies that for our purposes, rational 
maps are sufficiently general.
\begin{theorem}[Segal] 
\label{t:homotopyequiv}
$M_B^*$ and $Rat_B^*$ are homotopy equivalent up to $B$. \hfill $\square$
\end{theorem}

\noindent {\bf Proof of theorem \ref{t:FR}:}

First we show that the suspension ${\cal S}$ induces an surjective homomorphism 
${\cal S}_*: \pi_1(M_B^*) \to \pi_1(Q_B^*)$.
We only need to prove this  
for $B=0$ because Whitehead's theorem \ref{Whitehead} implies that all the 
fundamental groups $\pi_1(M_B^*)$ are isomorphic and similarly for 
$\pi_1(Q_B^*)$. It can be shown by the same argument as in 
\cite{Finkelstein:1966ft} that 
\begin{equation}
\pi_1(M_0^*) \cong \pi_3(S^2) \cong {\mathbb Z}.
\end{equation}
Recall that $\pi_1(Q_0^*) \cong \pi_4(S^3)$. Then the Freudenthal
suspension theorem \ref{t:suspension} implies that ${\cal S}_*$ is a
surjective homomorphism. 

It follows from Segal's theorem that this result also 
holds for rational maps as long as $B>0$. Namely, we have the following 
homomorphisms
\begin{equation}
\label{epimorphism}
\pi_1(Rat_B^*) \to \pi_1(M_B^*) \to \pi_1(Q_B^*),
\end{equation}
where the first map is an isomorphism because of theorem 
\ref{t:homotopyequiv} and the second map is a surjective homomorphism. \hfill 
$\square$

\subsection{The Fundamental Group of Rational Maps}
\label{Fundamental}

In this section we show how to calculate the homotopy class of a loop
that is generated by a given symmetry.
Segal has proven the following theorem in \cite[Proposition 
6.4]{Segal:1979}:
\begin{theorem}[Segal]
$\pi_1(Rat^*_B) = {\mathbb Z}$ and is generated by the loop which moves one 
zero of a rational function once (clockwise) around one pole. \hfill 
$\square$
\end{theorem}
Note that the inclusion induces a surjective homomorphism of fundamental 
groups between based and unbased rational maps
\begin{equation}
\pi_1(Rat^*_B) \to \pi_1(Rat_B)
\end{equation}
such that the generator of $\pi_1(Rat^*_B)$ is mapped to the generator of 
$\pi_1(Rat_B)$. The fundamental group of unbased maps is $\pi_1(Rat_B) = 
{\mathbb Z}_{2B}$ \cite{Epshtein:1973,Segal:1979}.

Let $z_i$ be the zeros, and $p_j$ the poles of 
the rational map $R$. For a given loop $L$ in $Rat^*_B$ the zeros and 
poles move around in the complex plane as a function of a parameter $\phi \in 
[0,\Phi]$ such that $z_i(0) = z_i(\Phi) = z_i$ and $p_j(0) = p_j(\Phi) = p_j$. 
Define the integral  
\begin{equation}
\label{def:N}
N(L) = \frac{i}{2 \pi} \sum\limits_{i,j=1}^{B}
\int\limits_0^{\Phi}
\frac{(z_i^\prime(\phi) - p_j^\prime(\phi)) ~{\rm d} \phi}
{(z_i(\phi) - p_j(\phi))}.
\end{equation}
where ${}^\prime$ denotes differentiation with respect to $\phi$.
Note that the denominators in the integrals do not vanish because poles 
and zeros cannot coalesce. 
\begin{lemma}
\label{l:iso}
$N(L)$ is a homotopy invariant and counts the number of times zeros 
move around poles. Therefore, $N(L)$ provides an isomorphism $\pi_1(Rat_B^*) 
\to {\mathbb Z}$.
\end{lemma}
\noindent {\bf Proof of lemma \ref{l:iso}:}
For a given loop, $L$, zeros and poles move around in the complex plane 
such that zeros move into zeros and poles move into poles. Therefore, a 
loop induces a permutation of poles and zeros. These permutations 
split up into cycles. Changing variables via $z(\phi) = z_i(\phi) - 
p_j(\phi)$ we obtain
\begin{equation}
N(L) = \frac{i}{2 \pi} \sum\limits_{k}
\oint\limits_{C_k(L)} \frac{{\rm d}z}{z},
\end{equation}
where the second sum runs over all the cycles and 
the integral is written as a contour integral.
The loop $L$ in the space
of rational maps induces one or more contour integrals over the contours 
$C_k(L)$.
The value of the integral is given according to Cauchy's theorem by 
the residue of the simple pole at $z=0$ and the number of times the 
$C_k(L)$ enclose this pole.

Consider now a loop ${\tilde L}$ which is homotopic to $L$. This induces a
set of contours ${\tilde C}_j$ where the number of cycles could have changed. 
Then $N({\tilde L}) - N(L)$ is a difference of contour integrals.
By adding and subtracting additional paths we can express 
$N({\tilde L}) - N(L)$ as a sum of closed contours $K_i(L,{\tilde L})$. 
However, 
because ${\tilde L}$  and $L$ are homotopic, there can be no residue inside 
$K_i(L,{\tilde L})$ as this would correspond to zeros coalescing with poles. 
Therefore, the integral over $K_i(L,{\tilde L)}$ vanishes by Cauchy's theorem 
and we have shown that $N({\tilde L}) = N(L)$, {\it i.e.} $N(L)$ is a homotopy 
invariant.

It is obvious from the properties of integration that $N(L_1 + L_2) = N(L_1) + 
N(L_2)$. Therefore, $N$ is a homomorphism. Furthermore, it is easy to check by 
direct calculation that moving a zero once clockwise around a pole gives 
rise to $N=1$. 
Therefore, $N$ provides an isomorphism between the fundamental 
group of based rational maps and the integers. \hfill $\square$

\begin{corollary}
\label{c:N}
The rational map ansatz induces an isomorphism $N \mod 2 \to \pi_1(Q_B^*)$.
\end{corollary}
{\bf Proof of corollary \ref{c:N}:}
Due to theorem \ref{t:FR} there is a surjective homomorphism 
$\pi_1(Rat_B^*) \to \pi_1(Q_B^*)$. According to lemma \ref{l:iso}, $N$ 
provides an isomorphism $\pi_1(Rat_B^*) \to {\mathbb Z}$. Since 
$\pi_1(Q_B^*) \cong {\mathbb Z}_2$ there is an isomorphism between 
$N \mod 2$ and $\pi_1(Q_B^*)$. \hfill $\square$\\

\subsection{Axially-Symmetric Skyrmions}
\label{Axial}

In the next two sections, we show how to calculate $N$ for a loop given by 
a combined 
rotation and isorotation. In general, the isospin action gives rise to a 
complicated movement of zeros and poles. However, for axially symmetric 
maps the isospin action can be calculated explicitly. 
Therefore, we consider axially symmetric maps first and then  
generalise the results to maps which are symmetric under discrete symmetries.

There are many possible $U(1) \times U(1)$ actions on $Rat_B^*$. However, for 
degree $B$ rational maps there is only a one parameter family of axially 
symmetric maps.

\begin{lemma}
\label{l:axial}
The most general axially symmetric map in $Rat_B^*$ for $B \neq 0$ is 
given by
\begin{equation}
\label{R_axial}
R(z) = \frac{z^B-b}{z^B+b},
\end{equation}
for $b \neq 0$.
\end{lemma} 

\noindent {\bf Proof of lemma \ref{l:axial}:}
It is easy to check that the only rotation that leaves the boundary condition 
$R(z=\infty) =1$ invariant is a rotation by an angle $\alpha$ around the 
$x_3$-axis,
\begin{equation}
\label{rot3}
z \to {\rm e}^{i \alpha} z.
\end{equation}
Similarly, the only isorotation that leaves the boundary conditions 
invariant is an isorotation by $\beta$ around the 
$X_1$-axis,\footnote{Here, $x_i$ denotes a Cartesian coordinate system of 
the domain, and $X_i$ is a local Cartesian coordinate system of the 
codomain. 
$R$ and $z$ are related to $X_i$ and $x_i$, respectively, by stereographic 
projection.} 
\begin{equation}
\label{iso1}
R \to \frac{\cos(\beta/2) R - i \sin(\beta/2)}
{-i \sin(\beta/2) R + \cos(\beta/2)}.  
\end{equation}
In the following, when no rotation or isorotation axis is mentioned we 
implicitly assume (\ref{rot3}) and (\ref{iso1}), respectively.

Under an infinitesimal rotation by  $\alpha$ followed by an infinitesimal 
isorotation by $\beta$ a rational map 
\begin{equation}
R(z) = \frac{p(z)}{q(z)}
\end{equation}
of degree $B \neq 0$ transforms into
\begin{equation}
\label{infsym}
{\tilde R}(z) = R(z) + \frac{i}{q(z)^2} \left(\frac{\beta}{2} \left(p(z)^2 
- q(z)^2 \right) + \alpha z \left(p^\prime(z) q(z) - p(z) q^\prime(z)
\right)
\right) + 
O(\alpha, \beta),
\end{equation}
where ${}^\prime$ denotes differentiation with respect to $z$. The map is 
invariant under a continuous symmetry generated by $\alpha$ and $\beta$ if and 
only if the term in big brackets vanishes. Assuming that $\beta = 0$ leads 
to
\begin{equation}
\alpha z R^\prime(z) = 0
\end{equation}
which implies that either $\alpha = 0$ so that there is no continuous symmetry 
or $R(z)$ is constant almost everywhere in contradiction to $B \neq 0$. 
Therefore, $\beta$ does not vanish.

As $R(z)$ has degree $B$ we can write the polynomials $p(z)$ and $q(z)$ as
\begin{eqnarray}
p(z) &=& \sum\limits_{k=0}^{B} a_k z^k, \\
q(z) &=& \sum\limits_{k=0}^{B} b_k z^k, 
\end{eqnarray}
and because of the boundary conditions we can set $a_B = b_B = 1$. 
The rational map is invariant, if the first order term in equation 
(\ref{infsym}) vanishes, {\it i.e.} all the coefficients of the polynomials are 
equal to zero. The coefficient of $z^n$ can be written as
\begin{equation}
\sum\limits_{k} a_k a_{n-k} - b_k b_{n-k} + \gamma (2k-n) (a_k b_{n-k} - a_{n-k} 
b_k),
\end{equation}
where $\gamma = \alpha / \beta$ and the sum runs from $k=0$ to $n$ if $n \leq B$ 
and from $k = n-B$ to $B$ if $n > B$. For $n=0$ we obtain $a_0^2 - b_0^2 = 0$.

\subsubsection*{Case: $a_0 = b_0$}

By induction let $a_i = b_i$ for $0 \leq i < n$. 
Using the coefficients with $n \leq B$ we obtain
\begin{equation}
\label{induction}
(a_n - b_n)(1 + n \gamma) = 0.
\end{equation}
For $1 + n \gamma \neq 0$ this implies by induction that $p(z) = q(z)$, 
{\it i.e.} 
$R(z) = 1$ in contradiction to the assumption that the degree of $R(z)$ is 
$B\neq 0$.

We can also use the coefficients for $n \geq B$ to  
perform an induction for decreasing 
$i$ by $a_i = b_i$ for $B \geq i > n-B$, noting that $a_B = b_B = 1$. This leads 
to
\begin{equation}
\label{inductiondown}
(a_{n-B} - b_{n-B})(1 + \gamma (n - 2B)) = 0.
\end{equation}
Setting $\gamma = -1/m$ and $m = n-B$ in (\ref{inductiondown}) we obtain
\begin{equation}
B(a_m - b_m) = 0,
\end{equation}
which implies by induction and equation (\ref{induction}) that
$a_i = b_i$ for all $0 \leq i \leq B$, {\it i.e.} $p(z) = q(z)$. This  is 
again a contradiction to $R(z) \neq 1$. Therefore, $a_0=b_0$ does not give any 
solutions.

\subsubsection*{Case: $a_0 = -b_0$}

By induction let $a_i = -b_i$ for $0 \leq i < n$. Using the coefficients 
with $n \leq B$ we obtain
\begin{equation}
(a_n + b_n)(1 - n \gamma) = 0.
\end{equation}
This leads to a contradiction for $n = B$ if $\gamma \neq 1/B$. Therefore, we 
have to set $\gamma = 1/B$.
Now, we can perform the same inductions as in the previous case. From equation 
(\ref{inductiondown}) we obtain
\begin{equation}
(a_{n-B} - b_{n-B})(1 - (2B - n)/B) = 0.
\end{equation}
This implies that
\begin{equation}
\begin{array}{c c c}
a_i = - b_i & {\rm and} & a_i = b_i,
\end{array}
\end{equation}
for $i = 1, \dots, B-1$, so that all such $a_i$ and $b_i$ vanish. Furthermore, 
$a_n = b_n = 1$ and $a_0 = - b_0 = - b$. This is a well-defined rational 
map of degree $B$ unless $b=0$. \hfill $\square$ \\

The zeros and poles of the rational map (\ref{R_axial}) are distributed
equidistantly on the unit circle in the $z$ plane. 
Under an isorotation (\ref{iso1}) by $2 \pi$ a pole
moves into a neighbouring pole and a zero moves into a neighbouring zero.
However, under a rotation by $2 \pi$ all zeros and poles move round 
once around the unit circle and end at their respective starting 
position. A combination of a rotation by $\alpha$ and an isorotation by 
$\beta$ is a symmetry of the axial map (\ref{R_axial}) if and only if 
\mbox{$B \alpha - \beta = 2 \pi k$} for $k \in {\mathbb Z}$.
We will denote a 
path generated by a rotation by $\alpha$ and an isorotation by $\beta$ by 
$L(\alpha, \beta)$.

\begin{lemma} 
\label{l:Naxial}
Given the axially symmetric map {\rm (\ref{R_axial})} of degree $B \neq 0$
and a loop generated by rotations by $\alpha$ and isorotations by $\beta$ 
such that \mbox{$B \alpha - \beta = 2 \pi k$} for $k \in {\mathbb Z}$. Then 
$N(L(\alpha,\beta))$ is given by 
\begin{equation}
\label{Naxial}
N(L(\alpha, \beta)) = \frac{B}{2 \pi} (B \alpha - \beta).
\end{equation}
\end{lemma}

\noindent {\bf Proof:}
Rotations around the $x_3$-axis are particularly easy to handle. Let 
$z_i$ and $p_j$ be the zeros and poles of a rational map $R$. Then $z \to 
{\rm e}^{i\phi} z$  changes the zeros and poles to ${\tilde z_i} = 
{\rm e}^{-i\phi}  z_i$ and  ${\tilde p}_i = {\rm e}^{-i\phi} p_i$. A 
rotation by $\alpha$ then induces a path $L(\alpha, 0)$. 
With equation (\ref{def:N}) we 
obtain 
\begin{equation}
N (L(\alpha, 0)) = \frac{\alpha}{2 \pi} B^2.
\end{equation}
Note that this is true for any rational map $R \in Rat_B^*$.

Now, let $z_n$ and $p_n$ be the zeros and poles of $R$ in 
(\ref{R_axial}). Then the zeros and poles induced by the 
isorotation are given by
\begin{eqnarray}
{\tilde z}_n &=& {\rm e}^{i \beta/B} z_n, \\
{\tilde p}_n &=& {\rm e}^{i \beta/B} p_n.
\end{eqnarray}
This induces a path $L(0, \beta)$. Then $N(L(0, \beta))$ is given 
by
\begin{equation}
N(L(0, \beta)) = -\frac{\beta}{2 \pi} B.
\end{equation}
For a given symmetry both $N(L(\alpha,0))$ and $N(L(0, \beta))$ need not 
be integers. However, since $L(\alpha, \beta) = L(\alpha, 0) + 
L(0, \beta)$ is a loop, the homotopy invariant $N(L(\alpha, \beta))$ 
is an integer and is given by equation (\ref{Naxial}).
This completes the proof of lemma \ref{l:Naxial}. \hfill $\square$\\

\subsection{General Formula for $N(L)$}

In this section we generalise the formula of lemma \ref{l:Naxial} to 
rational maps in $Rat_B$. 
First we discuss how to relate symmetric rational maps $R 
\in Rat_B$ to symmetric maps ${\tilde R} \in Rat_B^*$. Let $R \in 
Rat_B$ have the following symmetry
\begin{equation}
\label{Rsymmetry}
R(z) = {\tilde M} \left(R \left( M(z) \right) \right),
\end{equation}
where $M$ corresponds to a rotation by $\alpha$ around ${\bf n}$ and 
${\tilde M}$ is an isorotation by $\beta$ around ${\bf N}$. 
For $\alpha \neq 2 \pi k$ for $k \in {\mathbb Z}$, $M$ only leaves the 
points
\begin{equation}
\label{z+n}
z_{{\bf n}} = \frac{n_1 + i n_2}{1 + n_3} 
\end{equation}
and
\begin{equation}
\label{z-n}
z_{-{\bf n}} = - \frac{n_1 + i n_2}{1 - n_3}
\end{equation}
fixed. Similarly, ${\tilde M}$ only leaves $R_{\pm{\bf N}}$ fixed, where
$R_{\pm{\bf N}}$ is defined as in (\ref{z+n}) and (\ref{z-n}),
respectively.
Therefore, equation (\ref{Rsymmetry}) implies that $R(z_{-{\bf n}}) = 
R_{\pm {\bf N}}$. By redefining ${\bf N}$ and $\beta$ if necessary we set  
\begin{equation}
R(z_{-{\bf n}}) = R_{{\bf N}}.
\end{equation}
\begin{lemma}
\label{l:Rat}
There are transformations ${\tilde M}_{\tilde T}$ and $M_T$ such that the 
map 
\begin{equation}
{\tilde R} ({\tilde z}) = {\tilde M}_{\tilde T} (R 
(M_T^{-1} ({\tilde z})))
\end{equation}
is invariant under
\begin{equation}
\label{M1M3trafo}
{\tilde R}( {\tilde z}) = {\tilde M}_1 ({\tilde R} (M_3 ({\tilde z}))),
\end{equation}
where ${\tilde M}_1$ is an isorotation by $\beta$ around the $X_1$-axis 
and $M_3$ is a rotation by $\alpha$ around the $x_3$-axis. Furthermore, 
${\tilde R}(\infty) = 1$. 
${\tilde R}$ is unique up to rotations around $x_3$ 
and isorotations around $X_1$.
\end{lemma}

\noindent {\bf Proof:}
${M_T}$ is induced by the $SO(3)$ rotation 
\begin{equation}
T: {\bf n} \mapsto {\bf n}_3 = 
\left(
\begin{array}{c}
0 \\
0 \\
1 
\end{array}
\right).
\end{equation}
We can choose orthonormal vectors ${\bf n}$, ${\bf u}$, and ${\bf v}$
such that 
\begin{equation}
T^{-1} = T^t = ({\bf u}, {\bf v}, {\bf n}),
\end{equation}
and $\det T = 1$. Similarly, ${\tilde M}_{{\tilde T}}$ is induced by
\begin{equation}
{\tilde T}: {\bf N} \mapsto {\bf N}_1 = 
\left(
\begin{array}{c}
1 \\
0 \\
0 
\end{array}
\right).
\end{equation}
Again, we choose orthonormal vectors ${\bf N}$, ${\bf U}$, and ${\bf V}$, such 
that  
\begin{equation}
{\tilde T}^{-1} = ({\bf N}, {\bf U}, {\bf V}),
\end{equation}
and $\det {\tilde T} = 1$. There is a $U(1) \times U(1)$ family of choices how 
to define ${\bf u}$ and ${\bf v}$, and ${\bf U}$ and ${\bf V}$. This is 
generated by rotations and isorotations around the symmetry axis ${\bf n}$ and 
${\bf N}$. 
Now, we can express the rotations and isorotations as
\begin{equation}
\label{Mrot}
M(z) = M^{-1}_T(M_3(M_T(z)))
\end{equation}
and
\begin{equation}
\label{Miso}
{\tilde M}(R) = {\tilde M}^{-1}_{{\tilde T}}({\tilde M}_1({\tilde 
M}_{{\tilde T}}(R))),
\end{equation}
where $M_3$ is a rotation by $\alpha$ around the $x_3$-axis and ${\tilde M}_1$ 
is an isorotation by $\beta$ around the $X_1$-axis.
By inserting equations (\ref{Mrot}) and (\ref{Miso}) into equation 
(\ref{Rsymmetry}) and acting with ${\tilde M}_{{\tilde T}}$, we obtain
\begin{equation}
{\tilde M}_{{\tilde T}}(R(M^{-1}_T (M_T(z))))
=
{\tilde M}_1({\tilde M}_{{\tilde T}}(R(M^{-1}_T(M_3(M_T(z)))))).
\end{equation}
Setting ${\tilde z} = M_T(z)$ and ${\tilde R}({\tilde z}) = {\tilde M}_{{\tilde 
T}}(R(M^{-1}_T {\tilde z}))$ we have shown equation 
(\ref{M1M3trafo}).
Furthermore,
\begin{equation}
{\tilde R}(\infty) = {\tilde M}_{{\tilde T}}(R(M^{-1}_T (\infty)) 
= {\tilde M}_{{\tilde R}}(R(z_{-{\bf n}})) 
= {\tilde M}_{{\tilde T}}(R_{{\bf N}}) = 1. 
\end{equation}
${\tilde R}(z) $ is unique up to the above mentioned $U(1) \times U(1)$ family 
of choices which corresponds to rotations around $x_3$ and isorotations around 
$X_1$. \hfill $\square$ \\

In order to uniquely define the rotation angle $\alpha$ we make the following 
choice. A rotation is expressed as 
$\exp (-i \alpha {\bf n} \cdot {\bf J})$
where $\alpha \in [- 2 \pi, 2 \pi]$ and the sign of  
${\bf n}$ is chosen such that $n_3 > 0$ or $n_3 = 0$ and $n_2 > 0$ or $n_1 = 
1$. Similarly, $\beta \in [-2 \pi, 2 \pi]$ and ${\bf N}$ is given by
$R(z_{-{\bf n}})$.
With the above lemmata we prove the following theorem.

\begin{theorem}
\label{t:N}
The value of $N$ for a given symmetry of a rational map $R \in Rat_B$ only 
depends on the rotation angle $\alpha$ and the isorotation angle $\beta$, where 
the angles are defined such that $R(z_{-{\bf n}}) = R_{{\bf N}}$. It 
is given by $N = \frac{B}{2 \pi} (B \alpha - \beta)$.
\end{theorem}
{\bf Proof:}
Let $R_0$ be a based rational map that is symmetric under a rotation by 
$\alpha$ followed by an isorotation by $\beta$. If this generates axial 
symmetry then we can apply lemma \ref{l:Naxial}.
We will now consider the case that 
the symmetry is finite.\footnote{A finite dimensional compact Lie group 
cannot have discrete subgroups of infinite order.}
Then $\alpha$ and $\beta$ can be written as $\alpha = 2 \pi m/n$ and 
$\beta = 2 \pi k/l$ where $k,m \in {\mathbb Z}$ and $n, l \in {\mathbb N}$.

In order to compute $N(L)$ we use the fact that $N(L)$ is an isomorphism. Rather 
than calculating $N(L)$ for $\alpha$ and $\beta$ we calculate $N({\tilde L})$ 
for ${\tilde \alpha} = n l \alpha$ and ${\tilde \beta} = n l \beta$, {\it i.e.} 
$n l$ times the original loop $L$. Then  $N(L)$ is given by 
$N({\tilde L})/(nl)$.
The advantage of the new loop ${\tilde L}$ is that it contains complete 
$U(1) \times U(1)$ orbits of rotations and isorotations. Such 
orbits always generate closed loops.

Let $R_0$ be a based rational map. Since $Rat^*_B$ is connected there is a 
path $R_t$ from the original map 
$R_0$ to a map with axial symmetry denoted by $R_1$. Since rotations and 
isorotations preserve the degree and rotations $M_3$ around $x_3$ and 
isorotations ${\tilde M}_1$ around $X_1$ preserve the base point, $M_3$ 
and ${\tilde M}_1$ give rise to a homotopy of loops 
${\tilde L}_t$ starting at $R_t$ 
and generated by ${\tilde \alpha}$ and ${\tilde \beta}$. All the 
loops ${\tilde L}_t$ are well defined and closed, and ${\tilde L}_0 = {\tilde 
L}$ and ${\tilde L}_1$ is the loop for an axial symmetric map.  
We can apply the formula of lemma \ref{l:Naxial} to ${\tilde L}_1$ and obtain
\begin{eqnarray}
\nonumber
N(L) &=& \frac{N({\tilde L})}{nl} = \frac{1}{nl} \frac{B}{2 \pi} (B {\tilde 
\alpha} 
- {\tilde \beta}) \\
\nonumber
 &=& \frac{B}{2 \pi} (B \alpha - \beta). 
\end{eqnarray}

Now, let $R$ be an arbitrary rational map of degree $B$, {\it i.e.} $R \in 
Rat_B$. Let its symmetry be a rotation by $\alpha$ around ${\bf n}$ and an 
isorotation by $\beta$ around ${\bf N}$ such that $R(z_{-{\bf n}}) = R_{{\bf 
N}}$. Then we can use lemma \ref{l:Rat} to transform $R \in Rat_B$ into a map 
$R_0$ and calculate $N$ as above.  This transformation is unique up to a $U(1) 
\times U(1)$ family of choices which does not change the value of $N$. This 
completes the proof of theorem \ref{t:N}. \hfill $\square$ \\

Formula (\ref{def:N}) can also be used to calculate $N$ for a given symmetry 
numerically. We performed these calculations for the symmetries of Skyrmions
up to $B=10$ and confirmed the results of theorem \ref{t:N}.

We will now explore a few simple consequences of theorem \ref{t:N} and corollary 
\ref{c:N}. First note that 
the loop generated by a rotation (or isorotation) is homotopic to an 
element of the fundamental group of $Rat_B$, namely $a^k$ where $a$ is the 
generator of $Rat_B$ and $k$ is an integer. 
We know that $a^{2k} = 1$, therefore, either $k \equiv B \mod 2B$ or $k 
\equiv 0 \mod 2B$.

Consider first a $2 \pi$ rotation, {\it i.e.} $N=B^2$. If $B$ is even 
then
\begin{equation}
B^2 \equiv 0 \mod 2B
\end{equation}
and the loop is contractible (even in $Rat_B$). If $B$ is odd then
\begin{equation}
\begin{array}{c c c}
B \equiv 1 \mod 2 & {\rm implies\ that} &
B^2 \equiv B \mod 2B
\end{array}
\end{equation}
so that the loop is non-contractible and is homotopic to $a^B$.
Notice that for $2 \pi$ isorotations we obtain $N = B$. Therefore, rotations
by $2 \pi$ and isorotations by $2 \pi$ are homotopic in $Rat_B$ if and only if 
$B$ is odd.

With the aid of corollary \ref{c:N} we recover the result of Giulini that 
rotations by $2 \pi$ give rise to nontrivial loops if and only if $B$ is
odd, \cite{Giulini:1993gd}. The same is true for isorotations by $2 \pi$.
In the following section, we will consider nontrivial loops due to 
symmetries of rational maps.

\section{Results}
\label{Results}

In this section we use the results of the previous sections to construct the 
ground states and some excited states of the Skyrme model, which are 
compatible with the FR constraints. First we describe the construction, 
then we present our results in a table and compare them to the literature 
and experimental data.

\subsection{Construction of Ground States}
\label{Construction}

Battye and Sutcliffe have calculated the minimal energy rational 
maps for $B$ up to $22$, and 
these maps are unique up to arbitrary rotations and isorotations. Since 
only $B=1$ and $B=2$ have continuous symmetries and these cases are already 
extensively discussed in the literature, we restrict our attention to $B>2$. 
Only the following discrete symmetries occur empirically: 
Dihedral symmetry $D_n$ for $2 \leq n \leq 6$, tetrahedral symmetry $T$, 
octahedral symmetry $O$ and  icosahedral symmetry $Y$. All these symmetry 
groups can be generated by two generators. In order to calculate the 
ground state for a given baryon number $B$ we perform the following steps: 

\begin{enumerate}
\item Choose a representative $R$ of the minimal energy rational map.
\item Choose two generators of the symmetry group.
\item For each generator determine ${\bf n}$ and $\alpha$, and ${\bf N}$ 
and $\beta$ such that $R(z_{-{\bf n}}) = R_{{\bf N}}$.
\item Calculate $N$ $({\rm mod}\ 2)$ for each generator. These are the FR 
constraints.
\item Decompose rotations and isorotations into irreducible representations.
\item Starting with the $J=0$ and $I=0$ (or $J = \frac{1}{2}$ and 
$I = \frac{1}{2}$) state calculate which states are allowed, that 
is are consistent with the imposed symmetry, using group theory.
\end{enumerate}

In the following we describe how to calculate irreducible 
representations, see {\it e.g.} \cite{Hamermesh:1962}.
Let $D_{ij}(g)$  be a matrix representation of a finite group $G$ of order 
$|G|$. Then $\chi(g) = {\rm Tr}~ (D_{ij}(g))$ is its character. Any 
representation of $G$ can be decomposed into irreducible representations 
with character $\chi_i$, and the following orthogonality relation holds:
\begin{equation}
\frac{1}{|G|} \sum\limits_{g \in G} {\bar \chi}_i(g) \chi_j(g) = \delta_{ij}.
\end{equation}
The irreducible representations have been calculated for all the relevant 
groups, {\it e.g.} in the table in \cite{Koster:1963}, also see 
\cite{Dresselhaus:1996} for the icosahedral group. We use the notation 
of \cite{Hamermesh:1962} whereby one dimensional representations are labelled 
with an $A$ and two dimensional representations by $E$. In this context a 
${}^\prime$ means that it is the representation of the double group, that is a 
$2 \pi$ rotation gives rise to minus the identity. 

There is a simple formula for characters of $SU(2)$ representations of 
dimension $D$. For a rotation by an angle $\theta$ around a unit vector 
${\bf n}$ the character is given by

\begin{equation}
\label{character}
\chi\big( \exp \left( -i \theta\ {\bf n} \cdot {\bf J} \right) \big) 
= \frac{\sin \frac{D \theta}{2}}{\sin \frac{\theta}{2}}.
\end{equation}

There is a significant difference between rotations and isorotations. For 
even baryon numbers $B$ rotations by $2 \pi$ are contractible. 
We can define an $SU(2)$ action on the homogeneous coordinates $[p,q]$. 
If $B$ is odd, then a rotation by $2 \pi$ 
gives rise to $[-p,-q]$, whereas if $B$ is even, it gives $[p,q]$. 
Therefore, the $SU(2)$ representation of the symmetry group can 
always be mapped to an $SO(3)$ representation if $B$ is even.
This implies that we can choose the two dimensional 
irreducible representation $E_1^\prime$ of the double group to impose 
the symmetry.

For isorotations, the situation is different because $2 \pi$ isorotations 
are always noncontractible in the space of rational maps, 
so a $2 \pi$ isorotation
always gives $[-p,-q]$. If $B$ is odd, the isospin transformation 
corresponds to a double group representation $E_k^\prime$. 
However, if $B$ is even, it is given by a representation which is not 
in the double group.

Let $g$ be a rotation by $\alpha$ around ${\bf n}$ followed by an 
isorotation by $\beta$ around ${\bf N}$. Then we can calculate the FR 
constraints for the symmetry transformation, and we define
\begin{equation}
\chi_{FR}(g) = \left\{
\begin{array}{cl}
1 & {\rm if\ }  {\rm contractible}, \\
-1& {\rm otherwise}.
\end{array}
\right.
\end{equation}
$\chi_{FR}(g)$ forms a one dimensional representation of the symmetry 
group, which is identical to its character.

Let the wave function $\psi$ transform under a tensor 
product of rotations and isorotations, namely
the $2J+1$ dimensional representation $J$ 
and the $2I+1$ dimensional representation $I$. As shown in the 
previous section, $I$ and $J$ are integers if $B$ is even and 
half-integers if $B$ is odd. If a Skyrmion is invariant under a symmetry 
group then equation (\ref{FR}) imposes the following additional 
constraint on the wave function:
\begin{equation}
\exp\left( -i \alpha\ {\bf n} \cdot {\bf J} \right)
\exp\left( -i \beta\ {\bf N} \cdot {\bf I} \right)
\psi = \chi_{FR}(g) \psi.
\end{equation}
The character of a tensor product is given by the product of the 
characters. Therefore, the number $n$ of representations for given quantum 
numbers $J$ and $I$ that are compatible with the FR constraints is 
\begin{equation}
\label{chiFR}
n = \frac{1}{|G|} \sum \limits_{g \in G} 
{\bar \chi}_{rot,J}(g) {\bar \chi}_{iso,I}(g) \chi_{FR}(g).
\end{equation}

\begin{table}[!htb]
\begin{center}
\begin{tabular}
{|l|l|l|l|c|}
\hline
& & & & \\
$B$ & Symmetry & $N_{g_1}$ & $N_{g_2}$ & FR-Rep \\
& & & & \\
\hline
$1$ & $SO(3)$ & --- & --- & $A_1$ \\
\hline
$2$& $D_\infty$ & $N_{C_\infty} = 0$ & $N_{C_2} = 1$
& $A_2$ \\
\hline
$3$ & $T(E_1^\prime,E_1^\prime)$ & $N_{C_3} = 2$ 
& $N_{C_2} = 6$ & $A_1$ \\
\hline
$4$ & $O(E_1^\prime,E_1)$ & $N_{C_4} = 2$ 
& $N_{C_3} = 8$ & $A_1$ \\
\hline
$5$ & $D_2(E_1^\prime,E_1^\prime)$ 
& $N_{C_2} = 10$ & $N_{C_2^\prime} = 10$  & $A_1$ \\
$5^*$ & $O(E_1^\prime,E_1^\prime)$ & $N_{C_4} = 5$ & $N_{C_3} =10$ & $A_2$ \\
\hline
$6$ & $D_4(E_1^\prime,E_1)$ & $N_{C_4} = 12$ & 
$N_{C_2^\prime} = 15$ & $A_2$ \\
\hline
$7$ & $Y(E_1^\prime,E_2^\prime)$ & $N_{C_5} = 14$ & $N_{C_3} =14$ & $A_1$ 
\\
\hline
$8$ & $D_6(E_1^\prime, E_1) $ & $N_{C_6} = 8$ 
& $N_{C_2^\prime} = 28$ & $A_1$ \\
\hline
$9$ & $D_4(E_1^\prime, E_1^\prime)$ & $N_{C_4} = 18$ 
& $N_{C_2^\prime}= 36$ & $A_1$ \\
$9^*$ & $T(E_1^\prime,E_1^\prime)$ & $N_{C_3} = 24$ &
$N_{C_2} = 36$ & $A_1$ \\
\hline
$10$ & $D_4(E_1^\prime, E_1)$  & $N_{C_4} = 20$ & 
$N_{C_2^\prime} = 45$ & $A_2$ \\
$10^*$ & $D_3(E_1^\prime, E_1)$ & $N_{C_3} = 40$ & $N_{C_2} = 45$ & $A_2$ \\
\hline
$11$ & $D_3(E_1^\prime,E_1^\prime)$  & $N_{C_3} = 44$ 
& $N_{C_2^\prime} = 55$ & $A_2$ \\
$11^*$ & $Y(E_1^\prime, E_1^\prime)$ & $N_{C_5} = 22$ & $N_{C_3} = 44$ & 
$A_1$ \\
\hline
$12$ & $T(E_1^\prime,A_2\oplus A_3)$  &$N_{C_3} =40$ & $N_{C_2} = 72$ & $A_1$ \\
\hline
$13$ & $O(E_1^\prime,E_1^\prime)$  & $N_{C_4} = 39$ & $N_{C_3}= 52$ & $A_2$ \\
$13^*$ & $D_4(E_1^\prime, E_1^\prime)$ & $N_{C_4} = 39$ 
& $N_{C_2^\prime} = 78$ & $A_3$  
\\
\hline
$14$ & $D_2(E_1^\prime,A_1\oplus A_3)$  & $N_{C_2}=98$
& $N_{C_2^\prime} = 91$ & $A_3$ \\
\hline
$15$ & $T(E_1^\prime,E_1^\prime)$  & $N_{C_3} = 80$
& $N_{C_2} = 120$  & $A_1$ \\
\hline
$16$ & $D_2(E_1^\prime,A_1\oplus A_3)$  & $N_{C_2} = 128$  
& $N_{C_2^\prime} = 120$ & $A_1$ \\
$16^*$ & $D_3(E_1^\prime,E_1)$  & $N_{C_3} = 96$ 
& $N_{C_2} = 120$ & $A_1$ \\
\hline
$17$ & $Y(E_1^\prime,E_2^\prime)$ & $N_{C_5} = 68$ & $N_{C_3} = 102$ & 
$A_1$ 
\\
$17^*$ & $O(E_1^\prime,E_1^\prime)$ & $N_{C_4} = 68$ & $N_{C_3} = 102$ & $A_1$ 
\\
\hline
$18$ & $D_2(E_1^\prime,A_1 \oplus A_3)$  & $N_{C_2} = 162$ 
& $N_{C_2^\prime} = 153$ & $A_3$ \\
\hline
$19$ & $D_3(E_1^\prime, E_1^\prime)$  & $N_{C_3} = 114$ 
&$N_{C_2} = 190$ & $A_1$ \\
$19^*$ & $T(E_1^\prime,E_1^\prime)$ & $N_{C_3} = 114$ & $N_{C_2} = 190$ &
$A_1$ \\
\hline
$20$ & $D_6(E_1^\prime, E_1)$  & $N_{C_6} = 60$ 
&$N_{C_2}=190$ & $A_1$ \\
\hline
$21$ & $T(E_1^\prime,E_1^\prime)$  & $N_{C_3} =154$ 
& $N_{C_2} =210$& $A_1$ \\
\hline
$22$ & $D_5(E_1^\prime,E_1)$  & $N_{C_5}=88$ 
&$N_{C_2}=231$ & $A_2$ \\
$22^*$ & $D_3(E_1^\prime,E_1)$ & $N_{C_3} = 176$ & $N_{C_2} = 231$ & $A_2$ \\
\hline
\end{tabular}
\caption{This table shows the Finkelstein-Rubinstein constraints for 
$B=1$ to $22$. For more details see text.\label{GroupResults}}
\end{center}
\end{table}

In the following we describe table \ref{GroupResults}.
The first column is the baryon number $B$. In the second column, we 
display the symmetry of the rational map, and also the corresponding 
irreducible representations of rotations and isorotations. In the 
following two columns we show the value of $N$ for the two generators of the 
symmetry. We have chosen to display $N$ rather than $N \mod 2$ because it 
contains some more information about how the symmetries are imposed. Yet, 
it is $N \mod 2$ that implies the FR representation given in the last 
column.

For $B=1$ the FR constraints are trivial because an isorotation is 
equivalent to a rotation. Similarly, for $B=2$ a rotation by $\alpha$ 
corresponds to an isorotation by $2 \alpha$, so that $N$ vanishes for 
axial symmetry. Yet, the $C_2$ rotation gives a non-trival value for $N$.
For some values of $B$ there are additional "excited" states labelled 
by $B^*$. 
These are the particularly symmetric maps for $B=5$ and $11$ in 
Ref. \cite{Houghton:1998kg}, and also maps that are very close to minimal 
energy rational maps as mentioned in \cite{Battye:2001qn}. These excited 
states $B^*$ give an indication of how important the symmetry of a 
configuration is for determining its ground state.

\subsection{Ground States for $B=1, \dots, 22$}
\label{Groundstates}

\begin{table}[!htb]
\begin{center}
\begin{tabular}
{|l|l|l|l||c|l||c|}
\hline
& & & & & &\\
$B$ & 
$|J\rangle |I \rangle_{0}$ 
& $ | J\rangle | I\rangle_{1}$ 
& $ | J\rangle | I\rangle_{2}$ 
& Experiment &
$|J\rangle |I \rangle_{Exp.}$ 
& Match \\
& & & & & & \\
\hline
$1$ & 
$|\frac{1}{2} \rangle |\frac{1}{2} \rangle$ &
$|\frac{3}{2} \rangle |\frac{1}{2} \rangle$ &
$|\frac{1}{2} \rangle |\frac{3}{2} \rangle$ &
${}^1_1{\rm H}$ & 
$|\frac{1}{2} \rangle |\frac{1}{2} \rangle$ & $\blacksquare$ \\
\hline
$2$& 
$| 1 \rangle | 0 \rangle$ &
$| 3 \rangle | 0 \rangle$ &
$| 0 \rangle | 1 \rangle$ &
${}^2_1 {\rm H}$  & 
$| 1 \rangle | 0 \rangle$ & $\blacksquare$ \\
\hline
$3$ & 
$| \frac{1}{2} \rangle | \frac{1}{2} \rangle$ &
$| \frac{5}{2} \rangle | \frac{1}{2} \rangle$ &
$| \frac{3}{2} \rangle | \frac{3}{2} \rangle$ & 
${}^3_2 {\rm He}$ &
$| \frac{1}{2} \rangle | \frac{1}{2} \rangle$ & $\blacksquare$ \\
\hline
$4$ & 
$| 0 \rangle | 0 \rangle$ &
$| 4 \rangle | 0 \rangle$ &
$| 0 \rangle | 1 \rangle$ & ${}^4_2{\rm He}$ & 
$| 0 \rangle | 0 \rangle$ & $\blacksquare$ \\
\hline
$5$ &
$| \frac{1}{2} \rangle | \frac{1}{2} \rangle$ & 
 $| \frac{3}{2} \rangle | \frac{1}{2} \rangle$ &
 $| \frac{1}{2} \rangle | \frac{3}{2} \rangle$ & 
$({}^5_2 {\rm He})$ & 
 $| \frac{3}{2} \rangle | \frac{1}{2} \rangle$ & 
$({}^5_2 {\rm He}^{*})$ \\
$5^*$ &
$| \frac{5}{2} \rangle | \frac{1}{2} \rangle$ & 
 $| \frac{7}{2} \rangle | \frac{1}{2} \rangle$ &
 $| \frac{3}{2} \rangle | \frac{3}{2} \rangle$ & 
$({}^5_2 {\rm He})$ & 
 $| \frac{3}{2} \rangle | \frac{1}{2} \rangle$ & 
$({}^5_2 {\rm He}^{*})$ \\
\hline
$6$ & 
$| 1 \rangle | 0 \rangle$ &
$| 3 \rangle | 0 \rangle$ &
$| 1 \rangle | 1 \rangle$ & 
${}^6_3 {\rm Li}$ & 
$| 1 \rangle | 0 \rangle$ & $\blacksquare$ \\
\hline
$7$ &
$| \frac{7}{2}\rangle | \frac{1}{2} \rangle$ 
& $| \frac{13}{2} \rangle | \frac{1}{2} \rangle$
& $| \frac{3}{2} \rangle | \frac{3}{2} \rangle$ &
${}^7_3 {\rm Li}$ & 
 $| \frac{3}{2} \rangle | \frac{1}{2} \rangle$ 
& 
$({}^7_3 {\rm Li}^{**})$ \\
\hline
$8$ & 
$| 0 \rangle | 0 \rangle$ &
$| 2 \rangle | 0 \rangle$ &
$| 0 \rangle | 1  \rangle$ & $({}^8_4 {\rm Be})$ & 
$| 0 \rangle | 0 \rangle$ & $\blacksquare$ \\
\hline
$9$ &
$| \frac{1}{2} \rangle | \frac{1}{2} \rangle$&
$| \frac{3}{2} \rangle | \frac{1}{2} \rangle$ &
$| \frac{1}{2} \rangle | \frac{3}{2} \rangle$ & 
${}^9_4 {\rm Be}$ & 
$| \frac{3}{2} \rangle | \frac{1}{2} \rangle$ &
$({}^9_4 {\rm Be}^{*})$ \\
$9^*$ &
$| \frac{1}{2} \rangle | \frac{1}{2} \rangle$ &
$| \frac{5}{2} \rangle | \frac{1}{2} \rangle$ &
$| \frac{3}{2} \rangle | \frac{3}{2} \rangle$ & 
${}^9_4 {\rm Be}$ & 
$| \frac{3}{2} \rangle | \frac{1}{2} \rangle$ &
$({}^9_4 {\rm Be}^{*})$ \\
\hline
$10$ &
$| 1 \rangle | 0 \rangle$ &
 $| 3 \rangle | 0 \rangle$ &
 $| 1 \rangle | 1 \rangle$ & 
${}^{10}_5 {\rm B}$ & 
 $| 3 \rangle | 0 \rangle$ & 
$({}^{10}_5 {\rm B}^{*})$ \\
$10^*$ & 
$| 1 \rangle | 0 \rangle$ &
 $| 3 \rangle | 0 \rangle$ &
 $| 1 \rangle | 1 \rangle$ & 
${}^{10}_5 {\rm B}$ & 
 $| 3 \rangle | 0 \rangle$ & 
$({}^{10}_5 {\rm B}^{*})$ \\
\hline
$11$ &
$| \frac{1}{2}  \rangle | \frac{1}{2} \rangle$ &
$| \frac{3}{2}  \rangle | \frac{1}{2} \rangle$ &
$| \frac{1}{2}  \rangle | \frac{3}{2} \rangle$ &
${}^{11}_5 {\rm B}$ &
$| \frac{3}{2}  \rangle | \frac{1}{2} \rangle$ &
$({}^{11}_5 {\rm B}^{*})$ 
 \\
$11^*$ & 
$| \frac{1}{2}  \rangle | \frac{1}{2} \rangle$ &
$| \frac{11}{2}  \rangle | \frac{1}{2} \rangle$ &
$| \frac{3}{2}  \rangle | \frac{3}{2} \rangle$ &
${}^{11}_5 {\rm B}$ &
$| \frac{3}{2}  \rangle | \frac{1}{2} \rangle$ &
$({}^{11}_5 {\rm B}^{*})$  \\
\hline
$12$ & 
$| 0 \rangle | 0 \rangle$ &
$| 3 \rangle | 0 \rangle$ &
$| 0 \rangle | 1 \rangle$ & $ {}^{12}_6 {\rm C}$ & 
$| 0 \rangle | 0 \rangle$ & $\blacksquare$ \\
\hline
$13$ & 
$| \frac{5}{2} \rangle | \frac{1}{2} \rangle$ &
$| \frac{7}{2} \rangle | \frac{1}{2} \rangle$ &
$| \frac{3}{2} \rangle | \frac{3}{2} \rangle$ & 
${}^{13}_6 {\rm C}$& 
$| \frac{1}{2} \rangle | \frac{1}{2} \rangle$ & 
$({}^{13}_6 {\rm C}^{***})$ \\
$13^*$ & 
$| \frac{3}{2} \rangle | \frac{1}{2} \rangle$ &
$| \frac{5}{2} \rangle | \frac{1}{2} \rangle$ &
$| \frac{1}{2} \rangle | \frac{3}{2} \rangle$ & 
${}^{13}_6 {\rm C}$& 
$| \frac{1}{2} \rangle | \frac{1}{2} \rangle$ & 
$({}^{13}_6 {\rm C}^{**})$ \\
\hline
$14$ & 
 $| 1 \rangle | 0 \rangle$ &
 $| 2 \rangle | 0  \rangle$ &
 $| 0 \rangle | 1 \rangle$ & 
${}^{14}_7 {\rm N}$&
 $| 1 \rangle | 0 \rangle$ & $\blacksquare$ \\
\hline
$15$ &
$| \frac{1}{2} \rangle | \frac{1}{2} \rangle$ &
$| \frac{5}{2} \rangle | \frac{1}{2} \rangle$ &
$| \frac{3}{2} \rangle | \frac{3}{2} \rangle$ & 
${}^{15}_7 {\rm N}$ &
$| \frac{1}{2}  \rangle | \frac{1}{2} \rangle$ &  
$\blacksquare$\\
\hline
$16$ &
$| 0 \rangle | 0 \rangle$ &
$| 2 \rangle | 0 \rangle$ &
$| 0 \rangle | 1 \rangle$ & ${}^{16}_8 {\rm O}$ & 
$| 0 \rangle | 0 \rangle$ & $\blacksquare$ \\
$16^*$ &
$| 0 \rangle | 0 \rangle$ &
$| 2 \rangle | 0 \rangle$ &
$| 0 \rangle | 1 \rangle$ & ${}^{16}_8 {\rm O}$ & 
$| 0 \rangle | 0 \rangle$ & $\blacksquare$ \\
\hline
$17$ &
$| \frac{7}{2} \rangle | \frac{1}{2} \rangle$ &
$| \frac{13}{2} \rangle | \frac{1}{2} \rangle$ &
$| \frac{3}{2} \rangle | \frac{3}{2} \rangle$ & 
${}^{17}_8 {\rm O}$ & 
$| \frac{5}{2} \rangle | \frac{1}{2} \rangle$ &  
$({}^{17}_8 {\rm O}^{*(8)})$ \\
$17^*$ &
$| \frac{1}{2} \rangle | \frac{1}{2} \rangle$ &
$| \frac{7}{2} \rangle | \frac{1}{2} \rangle$ &
$| \frac{3}{2} \rangle | \frac{3}{2} \rangle$ & 
${}^{17}_8 {\rm O}$ & 
$| \frac{5}{2} \rangle | \frac{1}{2} \rangle$ &  
$({}^{17}_8 {\rm O}^{*})$ \\
\hline
$18$ & 
$| 1 \rangle | 0 \rangle$ &
$| 2 \rangle | 0 \rangle$ &
$| 0 \rangle | 1 \rangle$ & 
${}^{18}_8 {\rm O}$ & 
$| 0 \rangle | 1 \rangle$ & $({}^{18}_9 {\rm F})$ \\
\hline
$19$ & 
$| \frac{1}{2} \rangle | \frac{1}{2} \rangle$ &
$| \frac{3}{2} \rangle | \frac{1}{2} \rangle$ &
$| \frac{1}{2} \rangle | \frac{3}{2}\rangle$  & 
${}^{19}_9 {\rm F}$ &
$| \frac{1}{2} \rangle | \frac{1}{2} \rangle$ & 
$\blacksquare$\\
$19^*$ & 
$| \frac{1}{2} \rangle | \frac{1}{2} \rangle$ &
$| \frac{5}{2} \rangle | \frac{1}{2} \rangle$ &
$| \frac{3}{2} \rangle | \frac{3}{2} \rangle$ & 
${}^{19}_9 {\rm F}$ &
$| \frac{1}{2} \rangle | \frac{1}{2} \rangle$ & 
$\blacksquare$\\
\hline
$20$ & 
$| 0 \rangle | 0 \rangle$ &
$| 2 \rangle | 0 \rangle$ &
$| 0 \rangle | 1 \rangle$ & 
${}^{20}_{10} {\rm Ne}$ & 
$| 0 \rangle | 0 \rangle$ & $\blacksquare$ \\
\hline
$21$ &
$| \frac{1}{2} \rangle | \frac{1}{2} \rangle$ &
$| \frac{5}{2} \rangle | \frac{1}{2} \rangle$ &
$| \frac{3}{2} \rangle | \frac{3}{2} \rangle$ & 
${}^{21}_{10} {\rm Ne}$ & 
$| \frac{3}{2} \rangle | \frac{1}{2} \rangle$ & 
$({}^{21}_{10} {\rm Ne}^{*})$ \\
\hline
$22$ &
$| 1 \rangle | 0 \rangle$ &
$| 3 \rangle | 0 \rangle$ &
$| 1 \rangle | 1 \rangle$ & 
${}^{22}_{10} {\rm Ne}$ & 
$| 0 \rangle | 1 \rangle$ & 
$({}^{22}_{11}{\rm Na}^{*})$ \\
$22^*$  & 
$| 1 \rangle | 0 \rangle$ &
$| 3 \rangle | 0 \rangle$ &
$| 1 \rangle | 1 \rangle$ & 
${}^{22}_{10} {\rm Ne}$ & 
$| 0 \rangle | 1 \rangle$ & 
$({}^{22}_{11}{\rm Na}^{*})$ \\
\hline
\end{tabular}
\caption{These are the ground states and two excited 
states for $B=1$ to $22$. 
For further details see 
text.\label{Resulttable}} \end{center}
\end{table}

Table \ref{Resulttable} shows the ground states that have been 
calculated with our methods and compares our results to the experimental  
data in the table of isotopes \cite{tableofisotopes:1996}. It is worth 
emphasising that nuclei with the same value of $I$ and $J$ are degenerate 
in the Skyrme model, since it only models the strong interaction. 
In other words, only the number of nucleons matters, not whether 
they are protons or neutrons.
Table \ref{Resulttable} is organised as follows. The first column gives the 
baryon number $B$ corresponding to the number of nucleons in the 
atomic nuclei. 
In the following three columns we display the ground 
state and two excited states which have been obtained from the zero mode 
quantization of the Skyrme model. 
The ground state is the state with the lowest value of $I$, 
{\it i.e.} either $0$ or $\frac{1}{2}$, and the lowest value of $J$ which 
is compatible with the constraints. 
One excited state is given by the second 
lowest value of $J$. The other excited state is given by the second lowest 
value of $I$, {\it i.e.} $1$ or $\frac{3}{2}$, and the lowest value of $J$ 
that is compatible with the constraints. 
This calculation makes assumptions 
about the relative magnitudes of $U_{ij}$, $V_{ij}$ and $W_{ij}$ in 
equation (\ref{Lred}), see Sect. \ref{Zero-mode}.  
These integrals have to be evaluated explicitly using the numerical 
solutions, and this might change the relative order of the states. 
Such calculations will  be left for a future publication.

In the column ``Experiment'' we cite the ground state 
for a given baryon number and the following column gives the value for $J$ and 
$I$. For $B=21$ and $B=22$ the value of $I$ is not given in 
\cite{tableofisotopes:1996}. We make the reasonable assumption that 
$I=\frac{1}{2}$ and $I=1$, respectively.
The last column labelled ``Match'' has a black box when the Skyrme ground
state agrees with experiment and otherwise shows the lowest nuclear 
state with the given values of $I$ and $J$. Here $*$ is short for first 
excited state, $**$ is the second excited state, etc. Nuclei in brackets 
are unstable. For some excited state the value of $I$ is missing in 
\cite{tableofisotopes:1996}.
Here we assume that it agrees with the value for the ground state.
The values for $B=1$ and $B=2$ are taken from the literature, 
\cite{Adkins:1983ya, Braaten:1988cc}. Also note that the results agree 
with Carson \cite{Carson:1991yv} for $B=3$ and also 
with Irwin \cite{Irwin:1998bs} who considered $B=4$ to $9$.\footnote{The
only exception is that for $B=4$ we found an excited state 
$| 0 \rangle | 1 \rangle$ which is lower than the $| 2 \rangle | 1 
\rangle$ state predicted in \cite{Irwin:1998bs}.}

The results for even baryon number are promising. Our calculations of the 
ground state agree with experiment for all but three cases, namely $B = 
10$, $18$ and $22$. 
For $B=10$ calculations predict the state $| 1 \rangle | 0 \rangle$ 
rather 
than $| 3 \rangle | 0 \rangle$. It is difficult to see in our scheme 
why the  $| 1 \rangle  | 0 \rangle$ state should have higher energy 
than the $| 3 \rangle | 0 \rangle$ state.
For $B=18$ the experimental ground state is $| 0 \rangle | 1 \rangle$ 
rather than $| 1 \rangle | 0 \rangle$. 
This deserves further investigation, because the 
ordering of the states makes assumptions about the integrals 
(\ref{intU}) -- (\ref{intV}).
Finally for $B=22$ our calculations predict a $| 1 \rangle | 0 \rangle$ 
state whereas the experimental ground state is probably a $| 0 \rangle | 1 
\rangle$ state, which is incompatible with our results.
Note that the value of the binding energy per nucleon  has local peaks at
$B=4$, $8$, $12$, $16$ and $20$. All these ground states are predicted
correctly.

The results for odd baryon numbers are not as good.
Our calculations agree with experiment for $B=1$, $3$, $15$, and $19$ but 
the disagree for $B = 5$, $7$, $9$, $11$, $13$, $17$, and $21$.
It is worth mentioning that for $B=5$ there is no stable nucleus, and 
in general atomic nuclei are more stable for even $B$ than for odd $B$.
However, there are reasons to believe that the zero mode 
approximation is less reliable for odd $B$. 
For $B=1$ it has been argued by many authors 
that the correct energy to minimise is not the classical mass $M$ (given 
by $E$ in (\ref{Estandard})) but the rotationally improved energy 
\begin{equation}
{\tilde E} = M + \frac{1}{2 \Theta} I(I+1) + E_{\vecpi},
\end{equation}
where $\Theta$ is the moment of inertia and $E_{\vecpi}$ is the 
additional energy for including a pion mass term, see 
\cite{Rajaraman:1986ty}. A similar construction 
has to be implemented for $B>1$. This could deform the 
Skymion, change its symmetry and thereby change its ground state.
For even $B$ this does not play such a big role as $I=0$ in the 
ground state so that the additional term does not contribute, and
as a first approximation, the $J$ terms are assumed to be a smaller 
perturbation. For odd $B$, however, $I= \frac{1}{2}$ so that the term 
always has to be taken into account.\footnote{For a path integral 
derivation of the rotationally improved Skyrmion see Ref. 
\cite{Dorey:1994fk}.}

\section{Conclusion}

In this paper we showed that it is possible to calculate the FR 
constraints directly from the rational map ansatz. The key idea is to 
think of the rational map ansatz as a suspension. We proved that a loop 
in configuration space $Q_B^*$ is contractible if and only if it is 
homotopic to a suspension of a loop in $Rat_B$ which is 
an element of even order in $\pi_1(Rat_B)$. 
Even though the fundamental group of rational maps is more 
complicated than the fundamental group of configuration space it is 
nevertheless possible to derive a formula to calculate the 
homotopy class for loops generated by rotations 
and isorotations. It is worth emphasising that this formula is 
mathematically rigorous and there is no approximation involved. Therefore, 
the rational map ansatz is not only a good approximation to the minimal 
energy configurations, but it also captures important topology of the
configuration space.

In order to quantize Skyrmions as described in Sect. 
\ref{Zero-mode} it is important to have an approximate moduli space, and 
it is here that approximations come into play. In this paper we chose 
the zero mode approximation which only takes the rotational and 
isorotational degrees of freedom into account, because this is the 
simplest nontrivial application of our results about FR constraints.
In Sect. \ref{Results} we calculated the ground states of the Skyrme 
model for baryon numbers up to $B=22$. 
We found agreement with Irwin \cite{Irwin:1998bs} who calculated the FR 
constraints using an analogy with monopoles. Our results agree with 
experiments for all even baryon numbers apart from $B=10$, $18$ and $22$. 
It also appears that the Skyrme model works best for stable nuclei. In the 
odd baryon sector, the results are not as promising. Yet, there are
reasons why the zero mode approximation does not work as well for odd
$B$, as discussed at the end of the previous section.

In the following we will comment on future work.
In this paper we only relied on group theory making assumptions about 
the relative magnitude of integrals which involve the minimal energy 
configurations. 
With the numerical solutions in \cite{Battye:2001qn} it is possible to 
refine our results, verify the ground states, and calculate quantities 
such as mass, charge radii, and magnetic moments. By considering the 
reflection symmetries of Skyrmions the parity 
of the ground states can be calculated, \cite{Irwin:1998bs}. 
Finally, it is vital to go beyond the zero mode approximation. 
The problem of imposing FR constraints was solved in this paper, so the 
difficult question is how to find a suitable approximation to the Skyrmion 
moduli space. 

The first step is to consider rotationally improved Skyrmions. These could 
change the symmetries of the configurations and therefore, give rise to 
improved  ground states. These changes will be particularly significant 
for odd $B$ because in this case the contribution of isorotations to the 
energy cannot vanish.

Another possibility is to calculate the lowest vibrational modes of a 
Skyrmion and their frequencies \cite{Barnes:1997nc, Barnes:1997qa, 
Baskerville:1999kk, Baskerville:1999ve, Houghton:1999uq}. 
Then the excited states, and maybe even some ground 
states, are combinations of rotational and vibrational states. Another 
approach would be to construct a better approximation to the Skyrmion 
moduli space. 
For $B=2$ Manton has constructed a $12$ dimensional 
``unstable'' manifold \cite{Manton:1988ba}
which describes the configuration space of low energy configurations and 
can be thought of as a moduli space with a potential. Leese {\it et al.} 
found a $10$ dimensional submanifold which corresponds to Skyrmions 
in the attractive channel. In \cite{Leese:1995hb} the $B=2$ Skyrmion was 
quantized in this attractive channel approximation and the result was 
significantly better than in the zero mode approximation.
Note that the moduli space of monopoles is conjectured to be related to 
the moduli space of attractive channel Skyrmions. Therefore, 
it might be possible to use monopole fields for quantizing Skyrmions.

\section*{Acknowledgements}

The author is grateful to N S Manton for suggesting the problem and for 
many helpful suggestions. The author also wants to thank J M Speight for 
numerous fruitful discussions and continuous support.

\renewcommand{\baselinestretch}{1}
\begin{small}

\bibliography{/home/toco/sk211/bib/biblio}
\label{lastref}
\end{small}
\end{document}